\documentclass[usenatbib]{mnras}

\usepackage[T1]{fontenc}
\usepackage{ae,aecompl}

\usepackage{graphicx}	
\usepackage{amsmath}	
\usepackage{amssymb}	

\newcommand{\Msun}           {\,{\rm M}_\odot}

\newcommand{\rh}            {r_{\rm h}}
\newcommand{\rd}            {r_{\rm d}}
\newcommand{\Mstr}          {M_{\rm str}}

\newcommand{\gtot}          {g_{\rm tot}}
\newcommand{\gbar}          {g_{\rm bar}}

\newcommand{\gdm}           {g_{\rm dm}}

\title[MDAR in $\Lambda$CDM]{The origin of the mass
  discrepancy--acceleration relation in $\Lambda$CDM}

\author[J.F. Navarro et al.]{Julio F. Navarro$^{1}$\thanks{E-mail:jfn@uvic.ca},
  Alejandro Ben\'{i}tez-Llambay$^{2}$,
  Azadeh Fattahi$^{1}$, \newauthor 
 Carlos S. Frenk$^{2}$,
  Aaron D. Ludlow$^{2}$, 
  Kyle A. Oman$^{1}$,
  Matthieu Schaller$^{2}$, \newauthor
Tom Theuns$^2$.\\
$^{1}$Department of Physics and Astronomy, University of Victoria, Victoria, BC, Canada V8P 5C2\\
$^{2}$Institute for Computational Cosmology, Department of Physics, Durham University, South Road, Durham, DH1 3LE, UK\\
$^{3}$Leiden Observatory, Leiden University, PO Box 9513, 2300 RA Leiden, The Netherlands\\
}

\date{Accepted XXX. Received YYY; in original form ZZZ}

\pubyear{2016}

\begin{document}
\label{firstpage}
\pagerange{\pageref{firstpage}--\pageref{lastpage}}
\maketitle

\begin{abstract}
  We examine the origin of the mass discrepancy--radial acceleration
  relation (MDAR) of disk galaxies. This is a tight empirical
  correlation between the disk centripetal acceleration and that
  expected from the baryonic component. The MDAR holds for most radii
  probed by disk kinematic tracers, regardless of galaxy mass or
  surface brightness. The relation has two characteristic
  accelerations; $a_0$, above which all galaxies are baryon-dominated;
  and $a_{\rm min}$, an effective minimum aceleration probed by
  kinematic tracers in isolated galaxies. We use a simple model to
  show that these trends arise naturally in $\Lambda$CDM. This is
  because: (i) disk galaxies in $\Lambda$CDM form at the centre of
  dark matter haloes spanning a relatively narrow range of virial
  mass; (ii) cold dark matter halo acceleration profiles are
  self-similar and have a broad maximum at the centre, reaching values
  bracketed precisely by $a_{\rm min}$ and $a_0$ in that mass range;
  and (iii) halo mass and galaxy size scale relatively tightly with
  the baryonic mass of a galaxy in any successful $\Lambda$CDM galaxy
  formation model. Explaining the MDAR in $\Lambda$CDM does not
  require modifications to the cuspy inner mass profiles of dark
  halos, although these may help to understand the detailed rotation
  curves of some dwarf galaxies and the origin of extreme outliers
  from the main relation. The MDAR is just a reflection of the
  self-similar nature of cold dark matter haloes and of the physical scales
  introduced by the galaxy formation process.
\end{abstract}

\begin{keywords}
dark matter -- galaxies: kinematics and dynamics -- galaxies: structure
\end{keywords}



\section{Introduction}
\label{SecIntro}

The outer rotation curves of disk galaxies clearly deviate from
Newtonian predictions based on the gravitational attraction of their
gaseous and stellar components \citep{1978ApJ...225L.107R,Bosma1978}.
These deviations are usually ascribed to massive, spatially extended
dark matter haloes, a conclusion strongly supported by independent
lines of evidence, such as gravitational lensing of background objects
by galaxies and clusters, as well as by the structure of the Doppler
peaks in the cosmic microwave background, which suggests that most
matter in the Universe is in some non-baryonic form that interacts
little with radiation. A review of the topic may be found in
\citet{2005PhR...405..279B} and a summary of the latest parameters
inferred from cosmological surveys may be found in \citet{Planck2016}.

Although the evidence for dark matter seems on balance overwhelming, a
number of curious features in the kinematic evidence for dark matter
in disk galaxies have attracted attention over the years. These have
been argued to challenge the dark matter
interpretation of the data, and have motivated work on alternative
theories of gravity. Popular amongst them is the idea that
Newtonian gravity breaks down in the regime of `low acceleration'
($a<a_0\sim 10^{-10}\,{\rm m}\,{\rm s}^{-2}$) reached in the outskirts of galaxy
disks, as in the MOND scenario proposed by \citet{1983ApJ...270..371M}. 

A chief attraction of this idea is that disk rotation curves show {\it
  obvious} deviations from Newtonian predictions only in that regime,
regardless of other properties of the galaxy, such as mass, surface
brightness, or gas content
\citep[][]{1990A&ARv...2....1S}. Furthermore, the amount of dark
matter needed to explain, at a given radius, the observed rotation
velocity seems to correlate strongly with the enclosed baryonic mass,
to the extent that the full rotation curve of most disks may often be
predicted solely from the spatial distribution of baryons \citep[see,
e.g.,][and references
therein]{2006AIPC..822..253S,2015MNRAS.446..330W}.  This is an intriguing
result, which has at times been ascribed to a `conspiracy' between the
disk and the halo, but which has also strengthened alternative
theories such as MOND, where such correlations are thought to arise more
naturally.

These issues have been revisited recently by
\citet{2016arXiv160905917M} and \citet{2016AJ....152..157L} using a
compilation of late-type galaxy rotation curves and $3.6\,\mu{\rm m}$
Spitzer photometry, the band where uncertainties in the stellar
mass-to-light ratio are minimized \citep{2001ApJ...550..212B}. These
authors show that, for galaxies in their sample,
the disk centripetal acceleration, $\gtot(r)=V_{\rm circ}^2(r)/r$,
correlates strongly with that inferred from the spatial distribution
of the baryonic component, $\gbar(r)$, a relation termed `the
mass discrepancy-radial acceleration relation', or MDAR for short.

The MDAR indicates that baryons dominate in regions of high
acceleration; i.e., $\gtot\approx\gbar$ when
$\gtot>a_0$. In addition, few galaxies probe accelerations below
a well defined minimum value of $a_{\rm min} \sim 10^{-11}\,{\rm m}\,{\rm s}^{-2}$.
The latter point is further strengthened when adding to the
sample the ultra-faint satellites of the Milky Way, which include some of the most
dark matter-dominated and lowest-acceleration galaxies known \citep{2016arXiv161008981L}.

These results have renewed interest in the origin of the MDAR, and in
its theoretical interpretation. Although some have argued that the
MDAR is tantamount to a natural law that requires `new physics'
\citep[e.g.,][]{2012LRR....15...10F,2012PASA...29..395K,2015CaJPh..93..250M},
others have claimed that the MDAR arises as a consequence of the
scaling relations between the size and mass of galaxies and dark
haloes in the current paradigm of structure formation, $\Lambda$CDM
\citep{2016MNRAS.456L.127D,2016arXiv161006183K,2016arXiv161007663L,2016arXiv160701800D}.

It is clear from the current debate, however, that for the latter
interpretation to gain wide acceptance the reason for the existence of
characteristic accelerations such as $a_0$ and $a_{\rm min}$ in
disk kinematic data must be clearly identified. Our aim
is therefore to outline a simple argument for the origin of the
MDAR within the $\Lambda$CDM framework, including a compelling
motivation for its asymptotic behavior and for the characteristic
accelerations imprinted in it.

Our contribution extends earlier work, such as that of
\citet{2000ApJ...534..146V}, who used a semi-analytic model to show
that the MDAR may be reproduced in $\Lambda$CDM when galaxies are
constrained to match the Tully-Fisher relation, or that of
\citet{2002ApJ...569L..19K}, who used cosmological
arguments to motivate the origin of $a_0$. These arguments point to a
well-defined link between the `allowed' combinations of size, stellar
and total mass of galaxies and the narrow scatter of the MDAR, which
we develop further below.

\begin{figure*}
  \hspace{-0.2cm}
  \resizebox{17.cm}{!}{\includegraphics{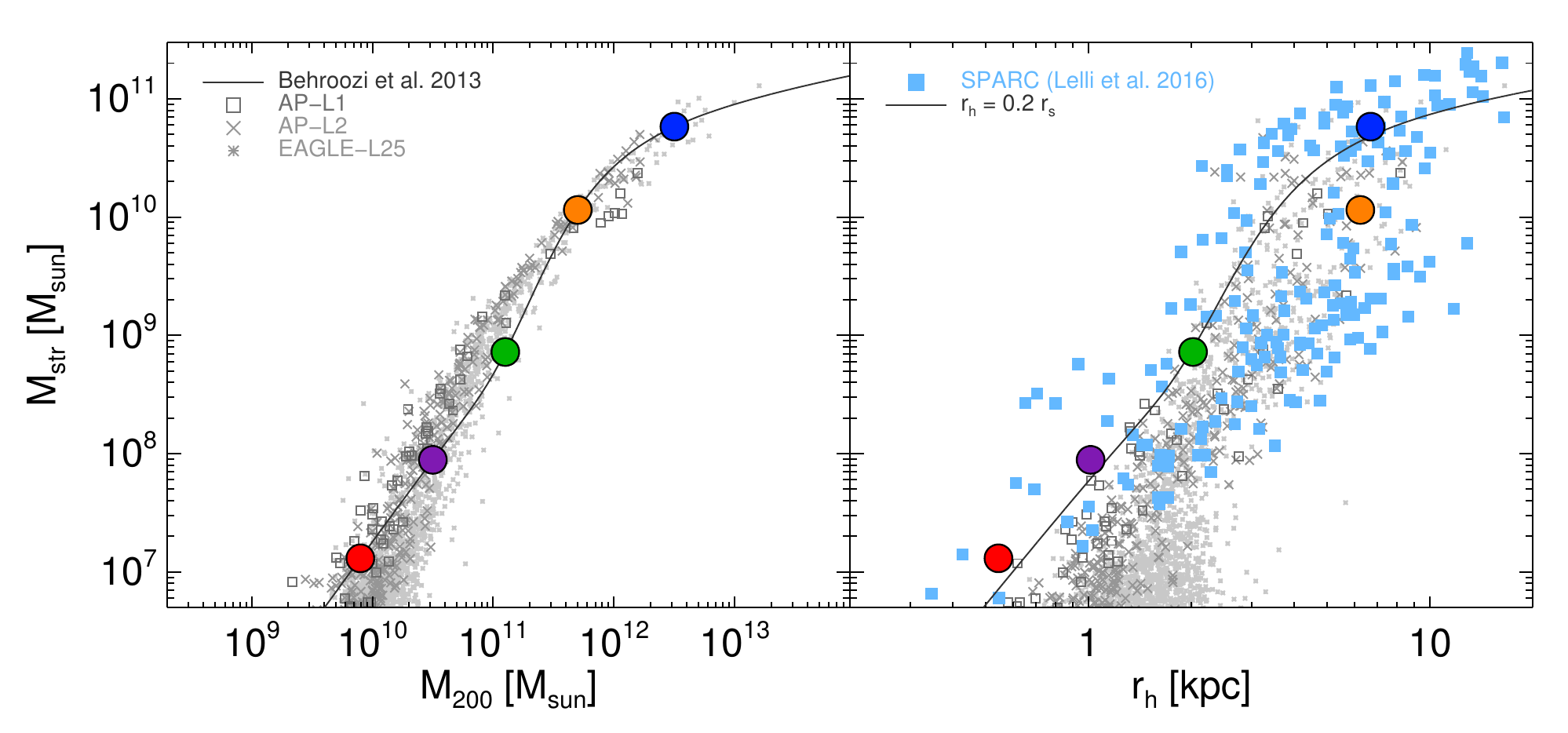}}\\%
  \caption{{\it Left:} Stellar mass vs. halo mass of five galaxies
    (solid circles), chosen to follow the abundance-matching relation
    of \citet{2013ApJ...770...57B}. Simulated galaxies from the
    {\small EAGLE} (Ref-L025N0752), {\small APOSTLE}-L1, and {\small
      APOSTLE}-L2 simulations are shown with asterisks, crosses, and
    open squares, respectively. The total stellar mass of the
    simulated galaxies ($M_{\rm str}$) corresponds to all bound star
    particles within a radius $r_{\rm gal}=0.15 \, r_{\rm 200}$. {\it
      Right:} Stellar mass vs. 3D stellar half-mass radii ($r_{\rm
      h}$). For the galaxy models we choose radii so that their
    characteristic accelerations follow the observed MDAR (see
    top-left panel of Fig.~\ref{FigModel}). The solid line indicates
    $r_{\rm h}=0.2 \, r_{\rm s}$, for reference. The stellar mass vs.
    $r_{\rm h}$ relation of model galaxies that match the MDAR is
    broadly consistent with the SPARC sample of galaxies. We assume
    $\rh=(4/3)R_{\rm eff}$ for the SPARC sample to account for
    projection effects. }
\label{FigMR}
\end{figure*}

\begin{figure*}
  \hspace{-0.2cm}
  \resizebox{17.8cm}{!}{\includegraphics{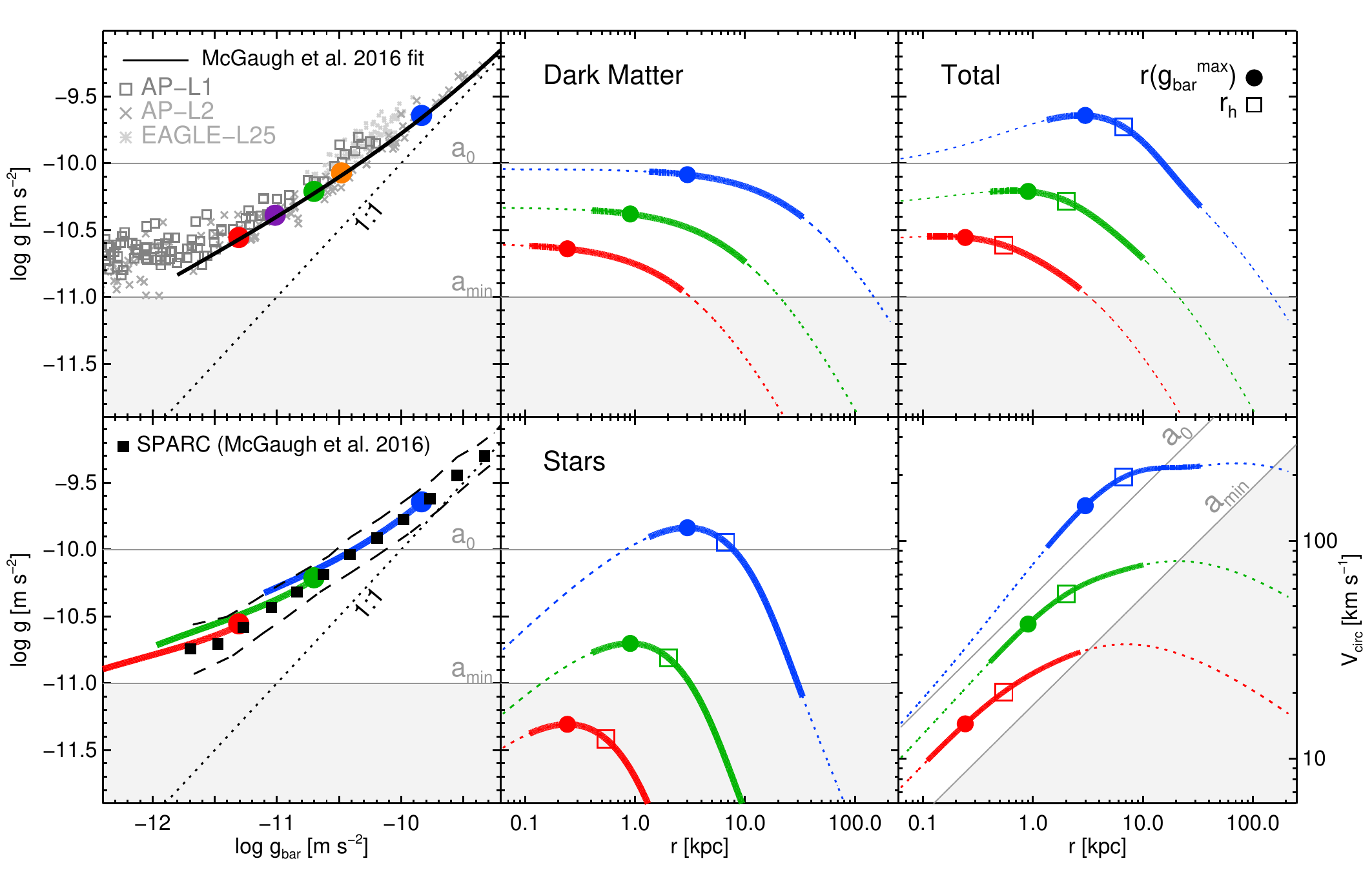}}\\%
  \caption{{\it Top-left:} Characteristic accelerations of the five
    model galaxies shown in Fig.~\ref{FigMR} (i.e., $g_{\rm bar}^{\rm
      max}$ and $g_{\rm tot}^{\rm max}$; filled circles) whose
    parameters have been chosen to match the MDAR of
    \citet{2016arXiv160905917M} (solid black line). Each simulated
    APOSTLE/EAGLE galaxy is shown once, evaluated at the stellar
    half-mass radius. {\it Bottom-left:} Same as top-left, but showing
    the radial acceleration profiles of three of the five model
    galaxies (for clarity). The radial range shown extends from $0.2\,
    r_{\rm h}$ to $5\, r_{\rm h}$ (thick lines). Filled circles are as
    in the top-left in all panels.The filled squares and dashed lines
    indicate the same MDAR (and its scatter) of
    \citet{2016arXiv160905917M}. {\it Top-middle:} Acceleration
    profiles, $g_{\rm dm}$, of the dark haloes of three of the models
    shown in the top-left panel. Dark matter haloes are assumed to
    follow NFW profiles, which have a broad maximum at the
    centre. {\it Bottom-middle:} Contribution of the baryons, $g_{\rm
      bar}$, to the acceleration profile in our galaxy models. For
    exponential disk stellar mass distributions, the acceleration
    reaches a characteristic maximum value of $g_{\rm bar}^{\rm max}$
    at $0.45\, r_{\rm h}$ (filled circles). Open squares indicate the
    stellar half-mass radius. {\it Top-right}: Same as middle panels,
    but for dark matter $+$ baryonic components. {\it Bottom-right:}
    Total circular velocity profiles of our models. See right-hand
    axis for units. In all panels the thick solid lines represent the
    radial range from $0.2\, r_{\rm h}$ to $5\, r_{\rm h}$, and the
    open squares mark $r_{\rm h}$. The shaded grey regions correspond
    to total acceleration below $a_{\rm min}=10^{-11}\,{\rm m}\,{\rm
      s}^{-2}$.  These regions are excluded for isolated galaxies,
    according to our simple model of galaxy formation in
    $\Lambda$CDM. See text for a full discussion.}
  \label{FigModel}
\end{figure*}

\section{The model}
\label{SecModel}

In $\Lambda$CDM, galaxies form at the centre of dark matter haloes
whose structural parameters and mass profiles are well understood
\citep[][hereafter, NFW]{1996ApJ...462..563N,1997ApJ...490..493N}. A large body of
numerical work has shown that cold dark matter haloes are well
approximated by NFW profiles, and may be characterized by two
parameters, usually expressed as a virial\footnote{Virial quantities
correspond to those of the sphere where the enclosed mean density is
$200$ times the critical density for closure,
$\rho_{\rm crit}=3H_0^2/8\pi G$, and are identified with a $200$
subscript.}  mass and a `concentration' parameter relating the
characteristic radius of a halo, $r_s$, to its virial radius,
$c=r_{200}/r_{s}$.  These two parameters are not independent.  The
$M_{200}(c)$ relation and its dependence on cosmological parameters is
now well understood \citep[see][and references
therein]{2014MNRAS.441..378L,2016MNRAS.460.1214L}, and therefore the
full mass profile of a
$\Lambda$CDM halo is known once its virial mass is specified.

In this context, the simplest galaxy formation model that may be used
to examine the MDAR requires the choice of a baryonic (stellar) mass
($M_{\rm str}$), a size and radial profile, as well as a way to relate
stellar mass to halo mass. The latter is probably the best understood
of those ingredients, given the strong constraint placed by the galaxy
stellar mass function on the halo mass--stellar mass relation in
$\Lambda$CDM. (We use `stellar' or `baryonic' indistinctly to refer to
the mass of the luminous component in this simple model.)

A simple, but reasonably accurate, parametrization of that relation is
provided by `abundance-matching' models, where galaxies are assigned
to dark matter haloes respecting their relative rankings by mass
\citep{Frenk1988,Vale2004,2010MNRAS.404.1111G,2013ApJ...770...57B,2013MNRAS.428.3121M}. The
solid line in the left panel of Fig.~\ref{FigMR} indicates the
relation derived by \citet{2013ApJ...770...57B} and compares it with
the results of the {\small EAGLE} and {\small APOSTLE} cosmological
hydrodynamical simulations\footnote{We show the results of the
  Ref-L025N0752 run of the {\small EAGLE} project, and L1 and L2 runs
  of {\small APOSTLE}. The {\small EAGLE} and {\small APOSTLE}-L2 runs
  have similar resolution, with gas particle mass of
  $\sim 10^5 \Msun$, while the {\small APOSTLE}-L1 runs have $10\times$
  better mass resolution, i.e. $\sim 10^4 \Msun$ per gas particle. All
runs use the same subgrid physical model.}
\citep{2015MNRAS.446..521S,2015MNRAS.450.1937C,2016MNRAS.457.1931S,2016MNRAS.457..844F}. These
simulations have been shown to match reasonably well the galaxy stellar
mass function over more than 4 decades in stellar mass; the shape of
disk galaxy rotation curves \citep{Schaller2015}; and the zero-point,
slope, and scatter of the Tully-Fisher relation
\citep{Ferrero2016}. Note that the stellar mass--halo mass relation is
rather steep at the faint end, implying that there is, broadly
speaking, an effective `minimum' halo mass required for a luminous
galaxy to form \citep[see, e.g.,][for further
discussion]{2016MNRAS.457.1931S,Benitez-Llambay2016}.

Galaxy sizes are known empirically to scale with stellar mass, as
shown, for example, by the SPARC sample of \citet[][filled squares in the right-hand panel of Fig.~\ref{FigMR}]{2016AJ....152..157L}. {\small
  EAGLE} and {\small APOSTLE} galaxies match the stellar half-mass
radius ($r_{\rm h}$) of SPARC galaxies fairly well, especially for
galaxies more massive than a few times $10^7\, M_\odot$. To first
order, $r_{\rm h}$ is well approximated by the relation
$r_{\rm h}=0.2 \, r_s$, as illustrated by the solid line in the right-hand panel of
Fig~\ref{FigMR}. These two relations show that in $\Lambda$CDM stellar
masses and sizes are inextricably linked to the masses and sizes of
their surrounding haloes.

The final choice of our model is a radial mass profile for the stellar (baryonic)
component of a galaxy, for which we adopt an exponential surface
density profile,
\begin{equation}
\Sigma_{\rm bar}(r)=\Sigma_0 e^{-r/\rd},
\label{EqExpDprof}
\end{equation}
where $r_{\rm d}= r_{\rm h}/1.678$ is the exponential scale radius and
the total disk mass is $\Mstr=2\pi
\Sigma_0 r_{\rm d}^2$.  
This is a good approximation to the
spatial distribution of stars in a typical galaxy disk.  

The total acceleration profile of the galaxy, $\gtot(r)$, may then be
calculated from the contributions of dark matter and stars, 
\begin{equation}
\gtot (r) = \gdm(r) +\gbar(r) =GM_{\rm dm}(<r)/r^2+V_{\rm bar}(r)^2/r,
\end{equation}
where $G$ is the gravitational constant; $M_{\rm dm}(<r)$ is the
enclosed mass of an NFW halo, corrected by a
factor\footnote{Cosmological parameters adopted throughout the paper
  are according to the Planck results $\Omega_{\rm m}=0.307$,
  $\Omega_{\rm \Lambda}=0.693$, $\Omega_{\rm bar}=0.04825$, and
  $H_{\rm 0}=67.77\, \rm{ km \, s^{-1}} Mpc^{-1}$
  \citep{2014A&A...571A..16P}.}
$(1-\Omega_{\rm bar}/\Omega_{\rm m})=0.84$ to account for the
universal baryon fraction, and $\gbar(r) = V_{\rm bar}(r)^2/r $ is
the contribution of the baryons to the centripetal acceleration.

Note that the dark matter contribution has a characteristic
acceleration, given by the central (maximum) value of an NFW profile:
$g_{\rm dm}^{\rm max}=\left[{c^2}/{(\ln(1+c)-c/(1+c))}\right]\left({V_{200}^2}/{2\,r_{200}}\right)$.
The baryons also have a well defined maximum acceleration,
$g_{\rm bar}^{\rm max}=0.286\,G\Mstr/r_{\rm d}^2$, which occurs at
$r_{\rm bar}^{\rm max}=0.747\,r_{\rm d}$. 

Each galaxy in our model therefore has a characteristic acceleration,
$g_{\rm tot}^{\rm max}$, given by the sum of these two values. Note
that $g_{\rm dm}$ and $g_{\rm bar}$ might peak at different radii so,
for simplicity, we shall adopt the total and baryonic accelerations at
$r_{\rm bar}^{\rm max}$ as the characteristic values for a model
galaxy. In practice,
$g_{\rm tot}^{\rm max}\approx g_{\rm tot}(r_{\rm bar}^{\rm max})$ so
this choice makes no difference to any of our results.

Finally, we have chosen to neglect here the response of the halo to the
assembly of the galaxy, mainly for simplicity but also because there
is still no overall consensus on the magnitude or even sign (i.e.,
contraction or expansion) of the effect.

\section{Results}
\label{SecRes}

\subsection{MDAR and scaling relations}
\label{SecMDAR-SR}

Disk rotation curves are best constrained around the baryonic half-mass
radius, where kinematic tracers are most abundant.  For our model to
be successful galaxies must therefore have characteristic
accelerations ($g_{\rm tot}^{\rm max}$ and $g_{\rm bar}^{\rm max}$)
that follow the MDAR. This condition places strong
constraints on the relation between galaxy stellar mass, size, and the
mass of its surrounding halo. We illustrate this in the top-left panel
of Fig.~\ref{FigModel}, where the filled circles correspond to five
example galaxies selected to follow the abundance matching relation
and to have radii so that their characteristic accelerations lie on
the MDAR. These examples span a range of nearly four decades
in stellar mass and more than one decade in radius. Their halo masses
are taken from the \citet{2013ApJ...770...57B} model, and their NFW
concentrations from the recent work of \citet{2016MNRAS.460.1214L}.

The example galaxies have radii quite consistent with the SPARC
mass-size relation, as may be seen in the right-hand panel of
Fig.~\ref{FigMR}. This shows that $\Lambda$CDM galaxies that follow
{\it simultaneously} the abundance-matching prescription (needed to
match the galaxy stellar mass function) and the empirical mass-size
relation can reproduce the observed MDAR without further adjustment.  

The MDAR thus results largely from the scaling relations linking the
size and mass of disk galaxies with the mass of their surrounding
halos. Indeed, the slight offset between the observed MDAR and that of
APOSTLE and EAGLE (top-left panel of Fig.~\ref{FigModel}) may be
traced to the slight and systematic devations of simulated galaxies from
both the abundance-matching and the empirical mass-size relations \citep[see
Fig. ~\ref{FigMR} and the discussion in][]{2016arXiv161007663L}.

\subsection{The origin of $a_0$ and $a_{\rm min}$}
\label{SecA0Amin}

The middle panels of Fig.~\ref{FigModel} explain the origin of the two
MDAR characteristic parameters; $a_0$ and $a_{\rm min}$, which, in
$\Lambda$CDM, result from the following considerations: (i) the
NFW acceleration profile has a well-defined maximum central value, and
declines very gradually with radius near the centre; (ii) the peak
acceleration varies by {\it only} a factor of $\sim 4$ for galaxies
that differ by a factor of $\sim 10^4$ in stellar mass; (iii) the peak
acceleration of the halo that hosts the most massive galaxy is very
nearly $a_0\approx 10^{-10}\,{\rm m}\,{\rm s}^{-2}$; and (iv) the
minimum acceleration $a_{\rm min}$ coincides with the NFW acceleration
at the outer edge (i.e., $r\sim 5\, r_{\rm h}$) of the faintest galaxy
in the examples.

Note that these results do not require any parameter tuning or
complicated galaxy formation model. They just rely on: (a) the NFW
mass profile shape, which has a well-defined, broad acceleration
maximum at the centre; (b) a reasonably tight correlation between
stellar mass and halo mass that satisfies the galaxy stellar mass
function; and (c) the limited radial range probed by luminous
kinematic tracers in galaxies.

Requisite (a) is a defining characteristic of
$\Lambda$CDM haloes, and one that does not necessarily hold for
alternative dark matter models. The peak accelerations in $\Lambda$CDM
haloes are determined by the cosmological parameters, which, unlike
more ad-hoc proposals like MOND, have {\it not} been tuned to fit
rotation curve data.

Condition (b) is a crucial outcome of any successful
$\Lambda$CDM galaxy formation model, and it is a result of the
baryon-driven energetic processes that regulate galaxy
formation. These processes
select a characteristic halo mass range outside of which galaxy formation
becomes extremely inefficient: at the centre of massive cluster-sized
halos, for example, where AGN feedback and long cooling times limit galaxy growth,
and in low-mass haloes, where the heating from cosmic reionization and
supernova feedback impose an effective minimum mass for halos that
host luminous galaxies. Galaxies in $\Lambda$CDM (and especially disks) 
thus form in a narrow range of halo virial velocity and an even narrower
range of central accelerations.

Finally, condition (c) is also important, since it predicts that extending
observations to radii well beyond the inner halo regions should lead
to systematic deviations from the MDAR.

The asymptotic behaviour of the $\gtot$--$\gbar$ relation can be
simply understood from the above discussion. Firstly, accelerations
larger than $a_0$ can {\it only} be reached in regions where baryons (which
may contract dissipatively and reach high densities/accelerations)
dominate. At accelerations greater than $a_0$, then, one expects
$\gtot\approx\gbar$, regardless of any other galaxy property. 

In regions where dark matter dominates, disk accelerations cannot drop
below $a_{\rm min}$, since that is roughly the minimum acceleration
traced in the observationally accessible range of the lowest mass
haloes that are effectively able to host a luminous isolated galaxy. The
model also predicts that dark matter-dominated dwarfs should have
acceleration profiles that vary weakly with radius, approaching a
constant $\gtot\sim a_{\rm min}$ at very low values of $\gbar$.

We emphasize that the latter conclusion applies only to isolated
`field' dwarfs, and {\it not} to satellite galaxies, which may see their
mass reduced by tidal stripping. Indeed, tidally stripped
satellites are expected to probe total acceleration values
significanty below $a_{\rm min}$, as in the case of the recently
discovered Milky Way satellite Crater II \citep{Caldwell2017}. The
relatively large size of this satellite and its extremely low velocity
dispersion are indicative of extremely low accelerations;
$g_{\rm tot} \sim 6\times 10^{-13}$ m~s$^{-2}$. Such extreme departure
from the minimum expected for field dwarfs in $\Lambda$CDM suggests
that Crater II must have been undergone large amounts of tidal
stripping, probably affecting both its dark matter and stellar
components. We plan to examine the consistency of this hypothesis with
observations of the Local Group satellite population in a separate
contribution.

\begin{figure*}
  \hspace{-0.2cm}
  \resizebox{17.8cm}{!}{\includegraphics{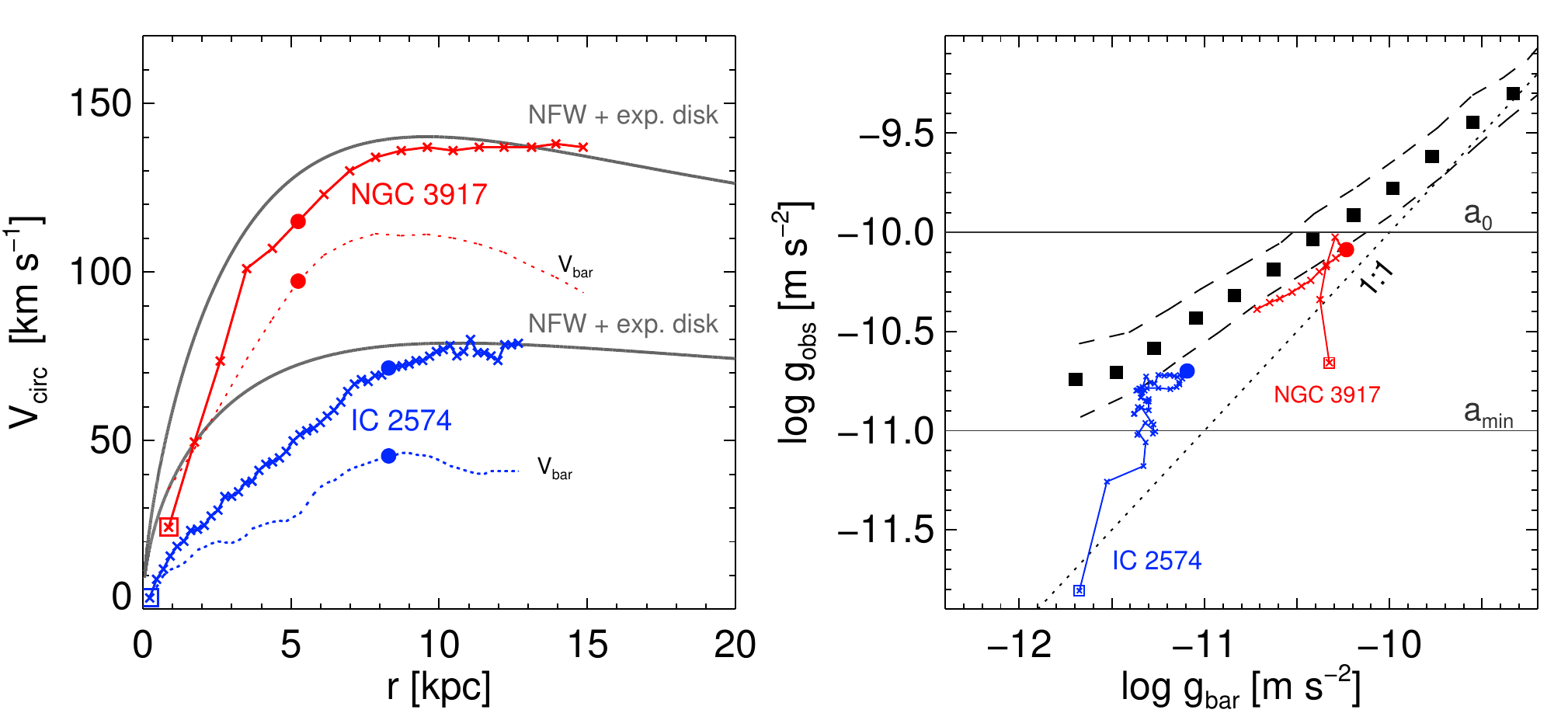}}\\%
  \caption{Illustration of the effect on the MDAR of alleged `cores'
    in the inner dark matter density profiles. {\it Left:} Rotation
    curves of two galaxies with large inner mass deficits. Lines
    connecting symbols are the inferred rotation curves; dotted lines
    are the contribution of the baryonic components. Filled circles
    indicate the location of the peak in the baryonic acceleration
    profile, $g_{\rm bar}(r)$. Thin grey lines are the result of our
    model, chosen to match the peak in the baryonic circular velocity
    profile, and the maximum rotation velocity of each galaxy. {\it
      Right:} The same two galaxies in the MDAR. Note that the inner
    regions, where the cores prevail, deviate systematically from the
    average MDAR. See text for further discussion. }
  \label{FigCores}
\end{figure*}

\subsection{MDAR and radial profiles}
\label{SecMDAR-RProf}

According to \citet{2016arXiv160905917M} and
\citet{2016AJ....152..157L}, the MDAR also appears to hold at various
radii of individual galaxies, an issue we address in the remaining
panels of Fig.~\ref{FigModel}. The top right panel shows the
centripetal acceleration profile, $g_{\rm tot}(r)$, of our example
galaxies (only three out of five are shown for clarity). The profiles
are shown in thick solid line type over the radial range typically
covered by kinematic tracers; from $20$~per~cent of $r_{\rm h}$ to $5
\times r_{\rm h}$. \citep[For reference, this corresponds to $\sim 0.7$--$18\,{\rm
  kpc}$  for a
galaxy like the Milky Way; see, e.g.,][]{2013ApJ...779..115B}. The filled circle indicates the
characteristic acceleration of the galaxy; i.e., the acceleration at
the radius where $g_{\rm bar}(r)$ peaks (see bottom middle panel of
Fig.~\ref{FigModel}).

Because the halo acceleration has a central maximum, and because the
peak baryonic acceleration occurs inside $r_{\rm h}$, neither the dark
matter nor the disk acceleration vary substantially over a wide radial
range, especially near the centre. This implies that the rotation
curve of an individual galaxy contributes many points to the MDAR just
around the characteristic value indicated by the solid circle in the
left panels of Fig.~\ref{FigModel}. This is in part responsible for the small
scatter reported for the MDAR, to an extent that depends on exactly
how the radial profile of individual galaxies is sampled, an issue to
which we shall return below.

Outside $r_{\rm h}$, the baryonic acceleration profile declines
rapidly with radius, extending the imprint of individual galaxies on
the MDAR to the left of each solid circle and following approximately
the average MDAR, as shown in the
bottom-left panel of Fig.~\ref{FigModel}. 

In systems where dark matter dominates (i.e., faint, low surface
brightness galaxies like the one identified in red) the total
acceleration changes little over the radial range where kinematic
tracers are present, explaining why the relation becomes nearly
horizontal at very small values of $g_{\rm bar}$.

On the other hand, in more massive, higher surface brightness systems
that are less baryon dominated (like the one identified in blue in
Fig.~\ref{FigModel}) the outer decline of the baryonic acceleration
profile is more pronounced, and leads to a steeper dependence of
$g_{\rm tot}$ with $g_{\rm bar}$ in the outer\footnote{Note that
  $g_{\rm tot}$ also declines towards the centre in systems where the
  disk dominates. This just reflects the importance of the disk in the
  overall potential and should not be confused with the presence of a
  constant density `core' in the dark matter, which may result in a
  similar trend in dark-matter dominated systems.}  regions. The
combination of these effects explains quite well the observed MDAR, as
shown in the bottom-left panel of Fig.~\ref{FigModel}.

\subsection{MDAR and dark matter `cores'}
\label{SecMDAR-Cores}

The previous discussion demonstrates that there is no need to
appeal to constant density `cores' in the inner dark matter
profile to explain the MDAR in $\Lambda$CDM, in agreement with the
conclusions of earlier work \citep[see, e.g.,][]{2016MNRAS.456L.127D,2016arXiv161006183K,2016arXiv161007663L,2016arXiv160701800D}.
Baryon-induced cores may be useful, however, to explain some outlier
points in the relation, such as those contributed by the inner
regions of galaxies whose rotation curves suggest the presence of a
core in the dark matter density profile---such cores are not
included in our simple model. Baryon-induced cores have also been
argued to improve agreement with the observed MDAR in the low-mass
galaxy regime, but the improvements refer to a small fraction of
outlier points and do not alter the main relation, at least for
a core-formation model like that of \citet[][see their fig.~4]{2016MNRAS.456L.127D}.

We illustrate the effect of cores in the MDAR by using data for two
galaxies whose rotation curves show an inner deficit of mass compared
with the predictions of $\Lambda$CDM models. As discussed by
\citet{2015MNRAS.452.3650O}, this deficit is a robust
characterization of the `core vs cusp' controversy, as shown in
Fig.~\ref{FigCores} for NGC 3917 \citep{2016AJ....152..157L} and IC
2574 \citep{2008AJ....136.2563W,2011AJ....141..193O}. The rotation
curve data are compared with the predictions of our simple model (grey
lines), after choosing disk and halo parameters to match the peak in
the baryonic circular velocity profile and the maximum observed
rotation velocity of each galaxy. Assuming that the rotation curves
faithfully trace the circular velocity profiles, the alleged `cores'
show up as a mismatch in the inner velocity profiles of model and
observation. These galaxies are two fairly extreme examples of alleged
cores, but are useful to illustrate the point.

As shown in the right-hand panel of Fig.~\ref{FigCores}, although the
characteristic acelerations of these two galaxies are not far from the
mean MDAR (filled circles), their inner regions show large systematic
deviations, even contributing a few points to the MDAR that dip
below the minimum acceleration $a_{\rm min}$ discussed in
Sec.~\ref{SecA0Amin}. Baryon-induced cores may help to explain these
outliers, but are not critical to the origin of the main MDAR trend
in $\Lambda$CDM, which is delineated by the relation between the characteristic
accelerations $g_{\rm tot}^{\rm max}$ and $g_{\rm bar}^{\rm max}$
discussed in Sec.~\ref{SecMDAR-SR}. 

We also note that the MDAR outliers arise from
acceleration estimates very near the galaxy centres, where rotation
velocities are low and where estimate uncertainties are magnified by
the non-negligible effects of non-circular motions and of the
`pressure' support provided by the finite gas velocity dispersion,
among other effects \citep[see, e.g.,][for some recent work on this topic]{Read2016,Pineda2017,Oman2017}.

\subsection{MDAR scatter}
\label{SecMDARScatter}

The discussion of the preceding subsection leads to the question of
why, if cores are as ubiquitous as is often claimed, the scatter in
the MDAR is as small as reported by \citet{2016arXiv160905917M} and
\citet{2016AJ....152..157L}. There are two reasons for this. One is
that cores as large and obvious as those of NGC 3917 and IC 2574 are
quite rare: indeed, most disk rotation curves only deviate mildly if
at all from $\Lambda$CDM expectations \citep[see,
e.g.,][]{2015MNRAS.452.3650O}.

The second is that the reported scatter is measured from an MDAR
constructed by sampling {\it linearly} in
radius the rotation curves of individual galaxies. This means that the
inner regions are de-emphasized in the average, which is dominated by
the large number of points that hover tightly around the
characteristic (peak) acceleration values of each galaxy.

This is shown in the right-hand panel of Fig.~\ref{FigCores} and is
particularly obvious in the case of NGC 3917: the inner regions
contribute only two points that deviate significantly from the average
MDAR. The scatter in the MDAR would probably be different if
each rotation curve was sampled logarithmically rather than linearly
in radius. In addition, individual points in a rotation curve are not
independent from each other when plotted as accelerations
(i.e., $g_{\rm bar}(r)$ is not a local measure but rather depends on the
whole baryonic mass profile), complicating the interpretation of the
scatter. 

This implies that a proper discussion of the MDAR scatter needs to
include a detailed consideration of the distribution of masses, radii,
and radial range sampling of galaxies in the SPARC sample. Although
this exercise is beyond the goals of this paper \citep[see][for a
recent attempt]{2016MNRAS.456L.127D}, we note that the MDAR
is a rather forgiving relation where even gross deviations from the
scalings assumed in our simple model translate into relatively small
changes to the predicted MDAR. This is a direct result of the narrow
range of central accelerations spanned by $\Lambda$CDM haloes that
host luminous disks, combined with the weak radial acceleration
gradient of the NFW profile. This issue has been discussed in more
detail in recent work \citep[see, e.g.,][]{Santos-Santos2016,2016arXiv161006183K,2016arXiv161007663L},
who used the results of direct cosmological simulations to discuss 
the MDAR scatter expected in $\Lambda$CDM.

\subsection{Deviations from MDAR}
\label{SecMDARDeviations}

We consider next the significance of deviations from the observed
MDAR. In $\Lambda$CDM the MDAR has no particular meaning, and one
would indeed expect systematic deviations in systems of much lower
or higher mass than halos that typically host field galaxies. Examples
include, at the low mass end, the haloes that host Ly-$\alpha$
absorbers at moderate redshift, and, at the massive end, rich galaxy
clusters, where MOND,
for example, fails to account for observations unless a dark mass
component is added \citep{Aguirre2001,Sanders2003}. 

In the context of our
discussion, we note that the acceleration at the centre of galaxy
clusters may exceed $a_0$. Indeed, the central NFW acceleration peaks
at $\sim 3\times 10^{-10}$ m s$^{-2}$ (i.e., three times higher than
$a_0$) for a cluster with $V_{200}\sim 1500$ km~s$^{-1}$, comparable to the
Coma cluster. Unfortunately, galaxy cluster centres are populated by
early-type galaxies, which are compact and massive enough to push the
observed accelerations to even higher values. The luminous regions of
these galaxies are expected therefore to populate the
$g_{\rm bar}\approx g_{\rm tot}$ region of the MDAR \citep[see,
e.g.,][]{2016arXiv161008981L}.

Alternatively, one might also expect strong deviations in very low
surface brightness galaxies, which trace the smallest\footnote{For
  practical purposes $\gbar\propto M_{\rm bar}(r)/r^2$ is just a proxy
  for enclosed surface brightness.} values of $\gbar$. If such
galaxies were to inhabit very massive haloes they would have high
$\gtot$ at low $\gbar$. Alternatively, if they were baryon-dominated,
they would have $\gtot \approx \gbar$ in the same regime,
deviating in both cases substantially from the mean MDAR
trend. Apparently such galaxies do not exist: very low surface brightness galaxies
form preferentially in low mass haloes and are dark matter
dominated. 

Finally, we note that $\Lambda$CDM predicts a high abundance of
very low mass halos where star formation has been fully prevented by cosmic
reionization. These halos, however, should still be filled with
(mostly ionized) gas,
and may be detectable in future H\,{\sc i} surveys \citep[see,
e.g.,][]{Benitez-Llambay2016}. Such systems should also
systematically deviate from the MDAR.

\section{Summary and Discussion}
\label{SecConc}

Recent work has highlighted the tight relation that links the radial
acceleration profile of galaxy disks, $\gtot(r)=V_{\rm circ}^2(r)/r$,
and that expected from their baryonic mass profile, $\gbar(r)$, for
disk galaxies spanning a vast range of stellar mass and surface
brightness. This mass discrepancy-radial acceleration relation (MDAR)
indicates that few, if any, known galaxies (a) probe accelerations
below a lower limit of $a_{\rm min}\sim 10^{-11}\,{\rm m}\,{\rm s}^{-2}$, or (b) are dark matter dominated at accelerations
exceeding $a_0\sim 10^{-10}\,{\rm m}\,{\rm s}^{-2}$.

We have used a simple model to show that the MDAR arises naturally in
$\Lambda$CDM. This is because (i) disk galaxies in $\Lambda$CDM form
at the centre of dark matter haloes spanning a relatively narrow range
of virial velocity ($30$--$300\,{\rm km}\,{\rm s}^{-1}$); (ii) dark
halo acceleration profiles are self-similar and have a broad maximum
at the centre, reaching values bracketed precisely by $a_{\rm min}$
and $a_0$ in that mass range; and (iii) halo mass and galaxy size
scale relatively tightly with the baryonic mass of a galaxy.

This implies that accelerations exceeding $a_0$ can {\it only} be
reached in regions that are dominated by baryons, explaining why
$\gbar\approx \gtot$ at high acceleration.  In addition, accelerations
cannot fall below $a_{\rm min}$ because of the effective minimum halo
mass needed to form a luminous galaxy, explaining why $\gtot\approx
a_{\rm min}$ at the very low values of $\gbar$ probed by dark
matter-dominated dwarf galaxies. 

Between those asymptotic limits, the MDAR follows from the
tight scaling between stellar mass and halo mass implied by the
baryonic physics that shapes the galaxy stellar mass function and from
the observed relation between stellar mass and size.  The
$\gbar$--$\gtot$ relation thus arises from the self-similar nature of
CDM haloes and of the physical scales introduced by the galaxy
formation process.

This also implies that isolated galaxies that deviate substantially
from the mean $\gbar$--$\gtot$ relation are difficult to account for
in $\Lambda$CDM. Examples include the dark matter `cores' inferred
for some galaxies from their slowly rising inner rotation curves,
which deviate from both the $\Lambda$CDM predictions {\it and} from the
average MDAR (see examples in Fig.~\ref{FigCores}).

If the inferred circular velocity curves for these galaxies are
correct, then they would invalidate {\it both} the views that the MDAR
encodes a `fundamental law' that goes beyond Newtonian gravity {\it
  and} that $\Lambda$CDM provides the framework for a correct theory
of structure formation. Galaxies such as these may thus reveal
potentially important modifications needed for both alternative models
of gravity and/or for $\Lambda$CDM.

A simpler alternative, however, is that the inferred circular velocity
curves in such galaxies are affected by substantially underestimated
systematic uncertainties. This is most likely the reason for the
outliers of the baryonic Tully-Fisher relation discussed as `missing
dark matter galaxies' by \citet{2016MNRAS.460.3610O}. However, it is
unclear whether such effects might be enough to bring galaxies like
IC~2574 into agreement with other galaxies, and with
$\Lambda$CDM. What is clear, however, is that such galaxies should be
thoroughly and carefully examined to establish whether they constitute
an insurmountable problem for $\Lambda$CDM or simply signal a
breakdown in the methods used to infer circular velocity curves from
gas velocity fields.

We end by identifying a population of galaxies where systematic
deviations from MDAR are to be expected. These are the low surface
brightness dwarf satellites of luminous galaxies, where tidal
stripping might reduce their dark matter content and velocity
dispersion while affecting little the size of the stellar component
\citep{Penarrubia2008}. Tidally-stripped dwarfs may thus dip below the
`minimum' acceleration ($a_{\rm min}$) expected for isolated
galaxies in $\Lambda$CDM. Given the strong dependence of the effects of
tides on orbital time and pericentric radius, one does not expect that
all satellites should be affected equally, leading to sizable scatter
in the $\gbar$--$\gtot$ relation at the very low surface brightness
end of the satellite population. There is tentative evidence that this
might indeed be the case, but a more detailed analysis is required to
gauge the role of tides on the structure of satellite galaxies.
Deviations from the MDAR may actually prove more revealing for our
understanding of galaxy formation than the relation itself.

\section{Acknowledgements}

We acknowledge the useful comments of the referee, which helped to
improve the presentation of these results. The research was supported
in part by the Science and Technology Facilities Council Consolidated
Grant (ST/F001166/1), and the European Research Council under the
European Union's Seventh Framework Programme (FP7/2007-2013)/ERC Grant
agreement 278594-GasAroundGalaxies. CSF acknowledges ERC Advanced
Grant 267291 COSMIWAY. This work used the DiRAC Data Centric system at
Durham University, operated by the Institute for Computational
Cosmology on behalf of the STFC DiRAC HPC Facility
(www.dirac.ac.uk). The DiRAC system was funded by BIS National
E-infrastructure capital grant ST/K00042X/1, STFC capital grants
ST/H008519/1 and ST/K00087X/1, STFC DiRAC Operations grant
ST/K003267/1 and Durham University. DiRAC is part of the National
E-Infrastructure. This research has made use of NASA's Astrophysics
Data System.

\bibliographystyle{mnras}
\bibliography{reflist}

\begin{thebibliography}{}
\makeatletter
\relax
\def\mn@urlcharsother{\let\do\@makeother \do\$\do\&\do\#\do\^\do\_\do\%\do\~}
\def\mn@doi{\begingroup\mn@urlcharsother \@ifnextchar [ {\mn@doi@}
  {\mn@doi@[]}}
\def\mn@doi@[#1]#2{\def\@tempa{#1}\ifx\@tempa\@empty \href
  {http://dx.doi.org/#2} {doi:#2}\else \href {http://dx.doi.org/#2} {#1}\fi
  \endgroup}
\def\mn@eprint#1#2{\mn@eprint@#1:#2::\@nil}
\def\mn@eprint@arXiv#1{\href {http://arxiv.org/abs/#1} {{\tt arXiv:#1}}}
\def\mn@eprint@dblp#1{\href {http://dblp.uni-trier.de/rec/bibtex/#1.xml}
  {dblp:#1}}
\def\mn@eprint@#1:#2:#3:#4\@nil{\def\@tempa {#1}\def\@tempb {#2}\def\@tempc
  {#3}\ifx \@tempc \@empty \let \@tempc \@tempb \let \@tempb \@tempa \fi \ifx
  \@tempb \@empty \def\@tempb {arXiv}\fi \@ifundefined
  {mn@eprint@\@tempb}{\@tempb:\@tempc}{\expandafter \expandafter \csname
  mn@eprint@\@tempb\endcsname \expandafter{\@tempc}}}

\bibitem[\protect\citeauthoryear{{Aguirre}, {Schaye}  \& {Quataert}}{{Aguirre}
  et~al.}{2001}]{Aguirre2001}
{Aguirre} A.,  {Schaye} J.,   {Quataert} E.,  2001, \mn@doi [\apj]
  {10.1086/323376}, \href {http://adsabs.harvard.edu/abs/2001ApJ...561..550A}
  {561, 550}

\bibitem[\protect\citeauthoryear{{Behroozi}, {Wechsler}  \&
  {Conroy}}{{Behroozi} et~al.}{2013}]{2013ApJ...770...57B}
{Behroozi} P.~S.,  {Wechsler} R.~H.,   {Conroy} C.,  2013, \mn@doi [\apj]
  {10.1088/0004-637X/770/1/57}, \href
  {http://adsabs.harvard.edu/abs/2013ApJ...770...57B} {770, 57}

\bibitem[\protect\citeauthoryear{{Bell} \& {de Jong}}{{Bell} \& {de
  Jong}}{2001}]{2001ApJ...550..212B}
{Bell} E.~F.,  {de Jong} R.~S.,  2001, \mn@doi [\apj] {10.1086/319728}, \href
  {http://adsabs.harvard.edu/abs/2001ApJ...550..212B} {550, 212}

\bibitem[\protect\citeauthoryear{{Ben{\'{\i}}tez-Llambay}
  et~al.,}{{Ben{\'{\i}}tez-Llambay} et~al.}{2016}]{Benitez-Llambay2016}
{Ben{\'{\i}}tez-Llambay} A.,  et~al., 2016, preprint, \href
  {http://adsabs.harvard.edu/abs/2016arXiv160901301B} {} (\mn@eprint {arXiv}
  {1609.01301})

\bibitem[\protect\citeauthoryear{{Bertone}, {Hooper}  \& {Silk}}{{Bertone}
  et~al.}{2005}]{2005PhR...405..279B}
{Bertone} G.,  {Hooper} D.,   {Silk} J.,  2005, \mn@doi [\physrep]
  {10.1016/j.physrep.2004.08.031}, \href
  {http://adsabs.harvard.edu/abs/2005PhR...405..279B} {405, 279}

\bibitem[\protect\citeauthoryear{{Bosma}}{{Bosma}}{1978}]{Bosma1978}
{Bosma} A.,  1978, PhD thesis, PhD Thesis, Groningen Univ., (1978)

\bibitem[\protect\citeauthoryear{{Bovy} \& {Rix}}{{Bovy} \&
  {Rix}}{2013}]{2013ApJ...779..115B}
{Bovy} J.,  {Rix} H.-W.,  2013, \mn@doi [\apj] {10.1088/0004-637X/779/2/115},
  \href {http://adsabs.harvard.edu/abs/2013ApJ...779..115B} {779, 115}

\bibitem[\protect\citeauthoryear{{Caldwell} et~al.,}{{Caldwell}
  et~al.}{2017}]{Caldwell2017}
{Caldwell} N.,  et~al., 2017, \mn@doi [\apj] {10.3847/1538-4357/aa688e}, \href
  {http://adsabs.harvard.edu/abs/2017ApJ...839...20C} {839, 20}

\bibitem[\protect\citeauthoryear{{Crain} et~al.,}{{Crain}
  et~al.}{2015}]{2015MNRAS.450.1937C}
{Crain} R.~A.,  et~al., 2015, \mn@doi [\mnras] {10.1093/mnras/stv725}, \href
  {http://adsabs.harvard.edu/abs/2015MNRAS.450.1937C} {450, 1937}

\bibitem[\protect\citeauthoryear{{Desmond}}{{Desmond}}{2016}]{2016arXiv160701800D}
{Desmond} H.,  2016, preprint, \href
  {http://adsabs.harvard.edu/abs/2016arXiv160701800D} {} (\mn@eprint {arXiv}
  {1607.01800})

\bibitem[\protect\citeauthoryear{{Di Cintio} \& {Lelli}}{{Di Cintio} \&
  {Lelli}}{2016}]{2016MNRAS.456L.127D}
{Di Cintio} A.,  {Lelli} F.,  2016, \mn@doi [\mnras] {10.1093/mnrasl/slv185},
  \href {http://adsabs.harvard.edu/abs/2016MNRAS.456L.127D} {456, L127}

\bibitem[\protect\citeauthoryear{{Famaey} \& {McGaugh}}{{Famaey} \&
  {McGaugh}}{2012}]{2012LRR....15...10F}
{Famaey} B.,  {McGaugh} S.~S.,  2012, \mn@doi [Living Reviews in Relativity]
  {10.12942/lrr-2012-10}, \href
  {http://adsabs.harvard.edu/abs/2012LRR....15...10F} {15, 10}

\bibitem[\protect\citeauthoryear{{Fattahi} et~al.,}{{Fattahi}
  et~al.}{2016}]{2016MNRAS.457..844F}
{Fattahi} A.,  et~al., 2016, \mn@doi [\mnras] {10.1093/mnras/stv2970}, \href
  {http://adsabs.harvard.edu/abs/2016MNRAS.457..844F} {457, 844}

\bibitem[\protect\citeauthoryear{{Ferrero} et~al.,}{{Ferrero}
  et~al.}{2016}]{Ferrero2016}
{Ferrero} I.,  et~al., 2016, preprint, \href
  {http://adsabs.harvard.edu/abs/2016arXiv160703100F} {} (\mn@eprint {arXiv}
  {1607.03100})

\bibitem[\protect\citeauthoryear{{Frenk}, {White}, {Davis}  \&
  {Efstathiou}}{{Frenk} et~al.}{1988}]{Frenk1988}
{Frenk} C.~S.,  {White} S.~D.~M.,  {Davis} M.,   {Efstathiou} G.,  1988,
  \mn@doi [\apj] {10.1086/166213}, \href
  {http://adsabs.harvard.edu/abs/1988ApJ...327..507F} {327, 507}

\bibitem[\protect\citeauthoryear{{Guo}, {White}, {Li}  \&
  {Boylan-Kolchin}}{{Guo} et~al.}{2010}]{2010MNRAS.404.1111G}
{Guo} Q.,  {White} S.,  {Li} C.,   {Boylan-Kolchin} M.,  2010, \mn@doi [\mnras]
  {10.1111/j.1365-2966.2010.16341.x}, \href
  {http://adsabs.harvard.edu/abs/2010MNRAS.404.1111G} {404, 1111}

\bibitem[\protect\citeauthoryear{{Kaplinghat} \& {Turner}}{{Kaplinghat} \&
  {Turner}}{2002}]{2002ApJ...569L..19K}
{Kaplinghat} M.,  {Turner} M.,  2002, \mn@doi [\apjl] {10.1086/340578}, \href
  {http://adsabs.harvard.edu/abs/2002ApJ...569L..19K} {569, L19}

\bibitem[\protect\citeauthoryear{{Keller} \& {Wadsley}}{{Keller} \&
  {Wadsley}}{2016}]{2016arXiv161006183K}
{Keller} B.~W.,  {Wadsley} J.~W.,  2016, preprint, \href
  {http://adsabs.harvard.edu/abs/2016arXiv161006183K} {} (\mn@eprint {arXiv}
  {1610.06183})

\bibitem[\protect\citeauthoryear{{Kroupa}}{{Kroupa}}{2012}]{2012PASA...29..395K}
{Kroupa} P.,  2012, \mn@doi [\pasa] {10.1071/AS12005}, \href
  {http://adsabs.harvard.edu/abs/2012PASA...29..395K} {29, 395}

\bibitem[\protect\citeauthoryear{{Lelli}, {McGaugh}, {Schombert}  \&
  {Pawlowski}}{{Lelli} et~al.}{2016a}]{2016arXiv161008981L}
{Lelli} F.,  {McGaugh} S.~S.,  {Schombert} J.~M.,   {Pawlowski} M.~S.,  2016a,
  preprint, \href {http://adsabs.harvard.edu/abs/2016arXiv161008981L} {}
  (\mn@eprint {arXiv} {1610.08981})

\bibitem[\protect\citeauthoryear{{Lelli}, {McGaugh}  \& {Schombert}}{{Lelli}
  et~al.}{2016b}]{2016AJ....152..157L}
{Lelli} F.,  {McGaugh} S.~S.,   {Schombert} J.~M.,  2016b, \mn@doi [\aj]
  {10.3847/0004-6256/152/6/157}, \href
  {http://adsabs.harvard.edu/abs/2016AJ....152..157L} {152, 157}

\bibitem[\protect\citeauthoryear{{Ludlow}, {Navarro}, {Angulo},
  {Boylan-Kolchin}, {Springel}, {Frenk}  \& {White}}{{Ludlow}
  et~al.}{2014}]{2014MNRAS.441..378L}
{Ludlow} A.~D.,  {Navarro} J.~F.,  {Angulo} R.~E.,  {Boylan-Kolchin} M.,
  {Springel} V.,  {Frenk} C.,   {White} S.~D.~M.,  2014, \mn@doi [\mnras]
  {10.1093/mnras/stu483}, \href
  {http://adsabs.harvard.edu/abs/2014MNRAS.441..378L} {441, 378}

\bibitem[\protect\citeauthoryear{{Ludlow} et~al.,}{{Ludlow}
  et~al.}{2016a}]{2016arXiv161007663L}
{Ludlow} A.~D.,  et~al., 2016a, preprint, \href
  {http://adsabs.harvard.edu/abs/2016arXiv161007663L} {} (\mn@eprint {arXiv}
  {1610.07663})

\bibitem[\protect\citeauthoryear{{Ludlow}, {Bose}, {Angulo}, {Wang},
  {Hellwing}, {Navarro}, {Cole}  \& {Frenk}}{{Ludlow}
  et~al.}{2016b}]{2016MNRAS.460.1214L}
{Ludlow} A.~D.,  {Bose} S.,  {Angulo} R.~E.,  {Wang} L.,  {Hellwing} W.~A.,
  {Navarro} J.~F.,  {Cole} S.,   {Frenk} C.~S.,  2016b, \mn@doi [\mnras]
  {10.1093/mnras/stw1046}, \href
  {http://adsabs.harvard.edu/abs/2016MNRAS.460.1214L} {460, 1214}

\bibitem[\protect\citeauthoryear{{McGaugh}}{{McGaugh}}{2015}]{2015CaJPh..93..250M}
{McGaugh} S.~S.,  2015, \mn@doi [Canadian Journal of Physics]
  {10.1139/cjp-2014-0203}, \href
  {http://adsabs.harvard.edu/abs/2015CaJPh..93..250M} {93, 250}

\bibitem[\protect\citeauthoryear{{McGaugh}, {Lelli}  \& {Schombert}}{{McGaugh}
  et~al.}{2016}]{2016arXiv160905917M}
{McGaugh} S.,  {Lelli} F.,   {Schombert} J.,  2016, preprint, \href
  {http://adsabs.harvard.edu/abs/2016arXiv160905917M} {} (\mn@eprint {arXiv}
  {1609.05917})

\bibitem[\protect\citeauthoryear{{Milgrom}}{{Milgrom}}{1983}]{1983ApJ...270..371M}
{Milgrom} M.,  1983, \mn@doi [\apj] {10.1086/161131}, \href
  {http://adsabs.harvard.edu/abs/1983ApJ...270..371M} {270, 371}

\bibitem[\protect\citeauthoryear{{Moster}, {Naab}  \& {White}}{{Moster}
  et~al.}{2013}]{2013MNRAS.428.3121M}
{Moster} B.~P.,  {Naab} T.,   {White} S.~D.~M.,  2013, \mn@doi [\mnras]
  {10.1093/mnras/sts261}, \href
  {http://adsabs.harvard.edu/abs/2013MNRAS.428.3121M} {428, 3121}

\bibitem[\protect\citeauthoryear{{Navarro}, {Frenk}  \& {White}}{{Navarro}
  et~al.}{1996}]{1996ApJ...462..563N}
{Navarro} J.~F.,  {Frenk} C.~S.,   {White} S.~D.~M.,  1996, \mn@doi [\apj]
  {10.1086/177173}, \href {http://adsabs.harvard.edu/abs/1996ApJ...462..563N}
  {462, 563}

\bibitem[\protect\citeauthoryear{{Navarro}, {Frenk}  \& {White}}{{Navarro}
  et~al.}{1997}]{1997ApJ...490..493N}
{Navarro} J.~F.,  {Frenk} C.~S.,   {White} S.~D.~M.,  1997, \apj, \href
  {http://adsabs.harvard.edu/abs/1997ApJ...490..493N} {490, 493}

\bibitem[\protect\citeauthoryear{{Oh}, {de Blok}, {Brinks}, {Walter}  \&
  {Kennicutt}}{{Oh} et~al.}{2011}]{2011AJ....141..193O}
{Oh} S.-H.,  {de Blok} W.~J.~G.,  {Brinks} E.,  {Walter} F.,   {Kennicutt} Jr.
  R.~C.,  2011, \mn@doi [\aj] {10.1088/0004-6256/141/6/193}, \href
  {http://adsabs.harvard.edu/abs/2011AJ....141..193O} {141, 193}

\bibitem[\protect\citeauthoryear{{Oman} et~al.,}{{Oman}
  et~al.}{2015}]{2015MNRAS.452.3650O}
{Oman} K.~A.,  et~al., 2015, \mn@doi [\mnras] {10.1093/mnras/stv1504}, \href
  {http://adsabs.harvard.edu/abs/2015MNRAS.452.3650O} {452, 3650}

\bibitem[\protect\citeauthoryear{{Oman}, {Navarro}, {Sales}, {Fattahi},
  {Frenk}, {Sawala}, {Schaller}  \& {White}}{{Oman}
  et~al.}{2016}]{2016MNRAS.460.3610O}
{Oman} K.~A.,  {Navarro} J.~F.,  {Sales} L.~V.,  {Fattahi} A.,  {Frenk} C.~S.,
  {Sawala} T.,  {Schaller} M.,   {White} S.~D.~M.,  2016, \mn@doi [\mnras]
  {10.1093/mnras/stw1251}, \href
  {http://adsabs.harvard.edu/abs/2016MNRAS.460.3610O} {460, 3610}

\bibitem[\protect\citeauthoryear{{Oman}, {Marasco}, {Navarro}, {Frenk},
  {Schaye}  \& {Ben{\'{\i}}tez-Llambay}}{{Oman} et~al.}{2017}]{Oman2017}
{Oman} K.~A.,  {Marasco} A.,  {Navarro} J.~F.,  {Frenk} C.~S.,  {Schaye} J.,
  {Ben{\'{\i}}tez-Llambay} A.,  2017, preprint, \href
  {http://adsabs.harvard.edu/abs/2017arXiv170607478O} {} (\mn@eprint {arXiv}
  {1706.07478})

\bibitem[\protect\citeauthoryear{{Pe{\~n}arrubia}, {Navarro}  \&
  {McConnachie}}{{Pe{\~n}arrubia} et~al.}{2008}]{Penarrubia2008}
{Pe{\~n}arrubia} J.,  {Navarro} J.~F.,   {McConnachie} A.~W.,  2008, \mn@doi
  [\apj] {10.1086/523686}, \href
  {http://adsabs.harvard.edu/abs/2008ApJ...673..226P} {673, 226}

\bibitem[\protect\citeauthoryear{{Pineda}, {Hayward}, {Springel}  \& {Mendes de
  Oliveira}}{{Pineda} et~al.}{2017}]{Pineda2017}
{Pineda} J.~C.~B.,  {Hayward} C.~C.,  {Springel} V.,   {Mendes de Oliveira} C.,
   2017, \mn@doi [\mnras] {10.1093/mnras/stw3004}, \href
  {http://adsabs.harvard.edu/abs/2017MNRAS.466...63P} {466, 63}

\bibitem[\protect\citeauthoryear{{Planck Collaboration} et~al.,}{{Planck
  Collaboration} et~al.}{2014}]{2014A&A...571A..16P}
{Planck Collaboration} et~al., 2014, \mn@doi [\aap]
  {10.1051/0004-6361/201321591}, \href
  {http://adsabs.harvard.edu/abs/2014A%26A...571A..16P} {571, A16}

\bibitem[\protect\citeauthoryear{{Planck Collaboration} et~al.,}{{Planck
  Collaboration} et~al.}{2016}]{Planck2016}
{Planck Collaboration} et~al., 2016, \mn@doi [\aap]
  {10.1051/0004-6361/201525830}, \href
  {http://adsabs.harvard.edu/abs/2016A%26A...594A..13P} {594, A13}

\bibitem[\protect\citeauthoryear{{Read}, {Iorio}, {Agertz}  \&
  {Fraternali}}{{Read} et~al.}{2016}]{Read2016}
{Read} J.~I.,  {Iorio} G.,  {Agertz} O.,   {Fraternali} F.,  2016, \mn@doi
  [\mnras] {10.1093/mnras/stw1876}, \href
  {http://adsabs.harvard.edu/abs/2016MNRAS.462.3628R} {462, 3628}

\bibitem[\protect\citeauthoryear{{Rubin}, {Thonnard}  \& {Ford}}{{Rubin}
  et~al.}{1978}]{1978ApJ...225L.107R}
{Rubin} V.~C.,  {Thonnard} N.,   {Ford} Jr. W.~K.,  1978, \mn@doi [\apjl]
  {10.1086/182804}, \href {http://adsabs.harvard.edu/abs/1978ApJ...225L.107R}
  {225, L107}

\bibitem[\protect\citeauthoryear{{Sanders}}{{Sanders}}{1990}]{1990A&ARv...2....1S}
{Sanders} R.~H.,  1990, \mn@doi [\aapr] {10.1007/BF00873540}, \href
  {http://adsabs.harvard.edu/abs/1990A%26ARv...2....1S} {2, 1}

\bibitem[\protect\citeauthoryear{{Sanders}}{{Sanders}}{2003}]{Sanders2003}
{Sanders} R.~H.,  2003, \mn@doi [\mnras] {10.1046/j.1365-8711.2003.06596.x},
  \href {http://adsabs.harvard.edu/abs/2003MNRAS.342..901S} {342, 901}

\bibitem[\protect\citeauthoryear{{Santos-Santos}, {Brook}, {Stinson}, {Di
  Cintio}, {Wadsley}, {Dom{\'{\i}}nguez-Tenreiro}, {Gottl{\"o}ber}  \&
  {Yepes}}{{Santos-Santos} et~al.}{2016}]{Santos-Santos2016}
{Santos-Santos} I.~M.,  {Brook} C.~B.,  {Stinson} G.,  {Di Cintio} A.,
  {Wadsley} J.,  {Dom{\'{\i}}nguez-Tenreiro} R.,  {Gottl{\"o}ber} S.,   {Yepes}
  G.,  2016, \mn@doi [\mnras] {10.1093/mnras/stv2335}, \href
  {http://adsabs.harvard.edu/abs/2016MNRAS.455..476S} {455, 476}

\bibitem[\protect\citeauthoryear{{Sawala} et~al.,}{{Sawala}
  et~al.}{2016}]{2016MNRAS.457.1931S}
{Sawala} T.,  et~al., 2016, \mn@doi [\mnras] {10.1093/mnras/stw145}, \href
  {http://adsabs.harvard.edu/abs/2016MNRAS.457.1931S} {457, 1931}

\bibitem[\protect\citeauthoryear{{Scarpa}}{{Scarpa}}{2006}]{2006AIPC..822..253S}
{Scarpa} R.,  2006, in {Lerner} E.~J.,  {Almeida} J.~B.,  eds,  American
  Institute of Physics Conference Series Vol. 822, First Crisis in Cosmology
  Conference. pp 253--265 (\mn@eprint {} {astro-ph/0601478}),
  \mn@doi{10.1063/1.2189141}

\bibitem[\protect\citeauthoryear{{Schaller} et~al.,}{{Schaller}
  et~al.}{2015}]{Schaller2015}
{Schaller} M.,  et~al., 2015, \mn@doi [\mnras] {10.1093/mnras/stv1067}, \href
  {http://adsabs.harvard.edu/abs/2015MNRAS.451.1247S} {451, 1247}

\bibitem[\protect\citeauthoryear{{Schaye} et~al.,}{{Schaye}
  et~al.}{2015}]{2015MNRAS.446..521S}
{Schaye} J.,  et~al., 2015, \mn@doi [\mnras] {10.1093/mnras/stu2058}, \href
  {http://adsabs.harvard.edu/abs/2015MNRAS.446..521S} {446, 521}

\bibitem[\protect\citeauthoryear{{Vale} \& {Ostriker}}{{Vale} \&
  {Ostriker}}{2004}]{Vale2004}
{Vale} A.,  {Ostriker} J.~P.,  2004, \mn@doi [\mnras]
  {10.1111/j.1365-2966.2004.08059.x}, \href
  {http://adsabs.harvard.edu/abs/2004MNRAS.353..189V} {353, 189}

\bibitem[\protect\citeauthoryear{{Walter}, {Brinks}, {de Blok}, {Bigiel},
  {Kennicutt}, {Thornley}  \& {Leroy}}{{Walter}
  et~al.}{2008}]{2008AJ....136.2563W}
{Walter} F.,  {Brinks} E.,  {de Blok} W.~J.~G.,  {Bigiel} F.,  {Kennicutt} Jr.
  R.~C.,  {Thornley} M.~D.,   {Leroy} A.,  2008, \mn@doi [\aj]
  {10.1088/0004-6256/136/6/2563}, \href
  {http://adsabs.harvard.edu/abs/2008AJ....136.2563W} {136, 2563}

\bibitem[\protect\citeauthoryear{{Wu} \& {Kroupa}}{{Wu} \&
  {Kroupa}}{2015}]{2015MNRAS.446..330W}
{Wu} X.,  {Kroupa} P.,  2015, \mn@doi [\mnras] {10.1093/mnras/stu2099}, \href
  {http://adsabs.harvard.edu/abs/2015MNRAS.446..330W} {446, 330}

\bibitem[\protect\citeauthoryear{{van den Bosch} \& {Dalcanton}}{{van den
  Bosch} \& {Dalcanton}}{2000}]{2000ApJ...534..146V}
{van den Bosch} F.~C.,  {Dalcanton} J.~J.,  2000, \mn@doi [\apj]
  {10.1086/308750}, \href {http://adsabs.harvard.edu/abs/2000ApJ...534..146V}
  {534, 146}

\makeatother
\end{thebibliography}

\label{lastpage}
\end{document}